\documentclass[onecolumn,showpacs,superscriptaddress,%
aps,pra,notitlepage,nofootinbib]{revtex4-1}

\usepackage{verbatim}
\usepackage{booktabs}
\usepackage{maybemath}

\begin{document}

\title{Binding Energy of Muonic Beryllium: \\
Perturbative versus All--Order Calculations}

\author{Shikha Rathi}
\affiliation{Physics Department, Technion--Israel Institute 
of Technology, Haifa 3200003, Israel}

\author{Ulrich D.~Jentschura}
\affiliation{Department of Physics and LAMOR, 
Missouri University of Science and Technology,
Rolla, Missouri 65409, USA}

\author{Paul Indelicato}
\affiliation{Laboratoire Kastler Brossel, 
Sorbonne Universit\'e, CNRS, ENS-PSL Research University, 
Coll\`ege de France, 4 Place Jussieu, F-75005 Paris, France}

\author{Ben Ohayon}
\affiliation{Physics Department, Technion--Israel Institute 
of Technology, Haifa 3200003, Israel}

\begin{abstract}
We compute the ground-state binding energy of muonic $^9$Be in two ways: 
first, the fully perturbative treatment of the 
nuclear-size effect often employed in light systems, and 
second, an approach that accounts for
the finite-nuclear-size to all orders
(and is inspired by calculations otherwise 
employed for heavy muonic ions).
The results are compared term by term and show that both approaches agree to better than one part-per-million of the total energy.
The objective of this work is twofold. The first is practical: to provide a parametrization that allows the extraction of the $^9$Be charge radius from recent and forthcoming experiments with high precision. The second is more conceptual: to act as a bridge between the community working on calculations for light systems and those focusing on heavy systems,
demonstrating that the fully relativistic approach otherwise chosen for heavy systems can be enhanced to cover theoretical predictions for all charge numbers.
\end{abstract}

\maketitle

\tableofcontents

\section{Introduction}

%% Preamable
Muonic atoms are simple, hydrogen-like systems that have a compact size approaching the nuclear scale. 
This makes their energy levels sensitive to both the structure of the nucleus~\cite{1967-def, 1969-mu,1974-Review, 1979-Friar,knecht2020study, antognini2022muonic, 2025-Q} and possible interactions beyond those included in the standard model, which are carried by a heavy ($\approx\,$MeV) new boson~\cite{2018-X17, 2022-Clara, beyer2025self, 2026-NP}.

The gross structure of muonic atoms lies in the x-ray/$\gamma$-ray regime. Its measurements started in the 1950's~\cite{1953-uX}, with activity peaking circa 1970~(see Appendix I in Ref.~\cite{1999-Angeli}).
At the turn of the millennium, the attention of the community shifted towards Lamb-shift measurements of the lightest muonic atoms using lasers~\cite{1999-uH}, culminating in precise measurements of muonic $^1$H~\cite{2013-Antognini}, $^2$H~\cite{2016-Pohl}, $^3$He~\cite{2025-3He}, and $^4$He~\cite{2021-Krauth}.

Just as the experimental techniques isolated the lightest muonic atoms from the rest, the theoretical approaches to calculate their energy levels have diverged.
For $Z=1$ and $Z=2$, the method of choice has predominantly been nonrelativistic quantum-electrodynamics (NRQED, see e.g.~\cite{2023-Theory} and references therein); a theoretical framework whose starting point is nonrelativistic dynamics and a point nucleus, which has the advantage of treating both bound particles on equal footing~\cite{2022-two-body, 2023-Adkins}.
For heavier systems, calculations are often based on a fully relativistic treatment of the orbiting muon in the field of an infinitely heavy nucleus with a realistic charge distribution (e.g.~\cite{1982-BR,2022-NP,2026-WK,2023-Ne}). 

%% Focus
In recent years, experimental studies of muonic-atom x-ray spectroscopy have seen renewed activity~\cite{2023-Ne, 2023-Microgram, 2024-SEVP,unger2024mmc,2024-MDPI,beyer2025modern, 2025-Ar}, with particular attention to the intermediate range $3\le Z<30$ where neither NRQED nor a fully-relativistic framework performs well. To make the most of these experimental efforts, a \textit{hybrid} theoretical framework applicable to all $Z$ has been introduced that treats the finite nuclear size non-perturbatively and consistently combines relativistic and nonrelativistic calculations~\cite{2026-Hybrid}. 

%% In this work
In this work, we assess the reliability of both the perturbative and the hybrid approach by computing various contributions to the same observable using each method and systematically comparing the outcomes. 
The chosen observable is the ground-state binding energy of muonic $^9$Be, without any electrons. 
In section~\ref{sec:main}, we give a detailed comparison, summarized in table~\ref{tab}. It is followed by section~\ref{sec:add}, in which further small corrections are discussed.
In section~\ref{sec:par}, we provide a parametrization of the main transition with the charge radius that is useful for extracting improved radii from experiments.

\section{Comparison of perturbative and all-order treatments of the finite nuclear size}\label{sec:main}
\begin{table}[bt]
    \centering
\begin{tabular}{l @{\hspace*{1cm}}rr@{\hspace*{1cm}}rr@{\hspace*{1cm}}r}
     \hline 
     \hline 
     Description & \multicolumn{2}{c}{All order }& \multicolumn{2}{c}{Perturbative}& Difference  \\
     & Eq.~\#{} & [eV] &  Eq.~\#{}s  & [eV] & [meV]\\
     \hline
    \midrule
    Dirac-Coulomb-Uehling  & \eqref{eq:DCU}       & $-45~072.59$ & \eqref{eq:Bohr}\,--\,\eqref{eq:eVP1-FNS2}       & $-45\,072.62$ & $30$ \\
    NR Recoil              & \eqref{eq:NR-Rec}    & $556.50$     & \eqref{eq:NMS}\,--\,\eqref{eq:FNS3-Rec}         & $556.51$ & $-3$ \\
    Källén-Sabry           & \eqref{eq:KS}        & $-$1.00      & \eqref{eq:KS-point}                             & $-$1.01 & 8 \\
    Muon Self-Energy       & \eqref{eq:SE1}       & 1.17         & \eqref{eq:SE1-point}\,--\,\eqref{eq:SE1-FNS}    & 1.18 & $-$11 \\
    Muonic Uehling         & \eqref{eq:uVP}       & $-0.05$      & \eqref{eq:uVPpoint}                             & $-0.05$ & 4 \\
    Hadronic Uehling       & \eqref{eq:hVP}       & $-0.03$      & \eqref{eq:hVP-point}                            & $-0.03$ & 3 \\
    NR Uehling Recoil      & \eqref{eq:eVP1-Rec}  & 2.89         & \eqref{eq:eVP1-Rec-NR}, \eqref{eq:eVP1-Rec-FNS} & 2.88 & 10\\
    NR Källén-Sabry Recoil & \eqref{eq:eVP2-Rec}  & 0.02         & \eqref{eq:eVP21-Rec-point}                      & 0.02 & 0\\
     \hline 
    Sum &  & $-$44\,513.08%2
    & & $-$44\,513.12 & 41 \\
     \hline 
    Relativistic Recoil & & & \eqref{eq:sal}& 0.04 &\\
    eVP SOPT Recoil & & & \eqref{Eq:eVP-SOPT-Rec}& 0.01 &\\
    Self-Energy Recoil & & & \eqref{eq:SE-Rec}& $-$0.05 &\\
    Self-Energy-eVP & & & \eqref{eq:SE-VP}& 0.01 &\\
    \midrule
     \hline 
    Sum &  & & & 0.01 & \\
     \hline 
     \hline 
    \bottomrule
    \end{tabular}
    \caption{Summary of calculations for the muonic $^9$Be ground-state energy.
    In both approaches, the non-recoil corrections are calculated 
    in the limit of an infinite nuclear mass.
    %before the reduced-mass correction is added and given separately for each relevant contribution
    %Here, we use the term ``recoil correction'' summarily for energy shifts caused by the finite nuclear mass. 
    NR stands for nonrelativistic. 
    }
    \label{tab}
\end{table}

\subsection{Definitions and Assumptions}

Our calculations are performed using a Gaussian charge distribution model, 
a root mean square (RMS) nuclear charge radius of $r_C=2.519\,$fm~\cite{1972-Be}, and a nuclear mass of $M=9.009989\,$amu~\cite{Wang_2021}.
Although $^9$Be has a spin of $3/2$, we do not consider its hyperfine structure here, and our calculations thus pertain to the hyperfine centroid energy.
Subleading relativistic-recoil corrections that depend on the nuclear spin are briefly discussed in subsection~\ref{sec:rel-rec}.
Moreover, corrections that are often packaged as stemming from \textit{nuclear structure}, 
namely those that account for nuclear (and nucleon) polarization and for a nuclear shape 
other than Gaussian, are not considered here, as the goal is to validate the QED calculations.

The small parameters associated with our calculations are the loop parameter $\alpha/\pi\approx0.2\%$; 
the binding parameter $Z\alpha\approx3\%$ (the nuclear charge number
is $Z=4$); the finite-size parameter 
\begin{equation}
    \rho\equiv r_C/a_0\approx4\% \,,
\end{equation}
for the effective Bohr radius $a_0=1/(Z\,\alpha\,m_\mu)$,
and the nuclear recoil parameter 
\begin{equation}
    \eta\equiv m_\mu/M\approx1\% \,,
\end{equation}
where we observe, for the reduced mass $m_r$,
the identity $m_r/m_\mu = 1/(1+\eta$).

\subsection{Dirac-Coulomb Solutions and Uehling Potential}\label{sub:DCU}

%% Hybrid approach %%
In the infinite nuclear mass limit, the hybrid approach is identical to the fully relativistic approach, where the main contribution is obtained by solving the Dirac-Coulomb equation for a muon bound to an infinitely-heavy nucleus of finite size. 
The leading QED effect, the one-loop electronic vacuum polarization 
(often denoted eVP$_{11}$, because it is of one-loop order and due to a single vertex involving the nucleus), 
is incorporated by adding the Uehling potential~\cite{1935-Uehling} to that of the nucleus.
The Dirac equation is then solved self-consistently. Here we used the multiconfiguration Dirac Fock and general matrix elements (MDFGME) code~\cite{2024-MCDFGME}, returning
\begin{equation}\label{eq:DCU}
E_\text{DCU}=-45\,072.59~\text{eV}.
\end{equation}
This result accounts for the Coulomb interaction and the leading electronic vacuum polarization to all orders in both $Z\alpha$ and $\rho$.

We next reproduce the value given in Eq.~(\ref{eq:DCU}) using a sequence of analytical and semi-analytical expressions expanded perturbatively in $\rho$ that are summed until the remaining unaccounted terms are small enough to be irrelevant for our target accuracy.
The starting point is the nonrelativistic Bohr energy for an infinitely heavy nucleus, which is (in a unit system where $\hbar = c = \epsilon_0 = 1$)
\begin{equation}\label{eq:Bohr}
E_0\equiv -\frac12 m_\mu (Z\alpha)^2=-45~011.61\,\text{eV}.
\end{equation}
Purely relativistic (but not recoil or those that pertain to the finite size)
corrections to Eq.~(\ref{eq:Bohr}), to all orders in $Z\alpha$, may be calculated by subtracting Eq.~(\ref{eq:Bohr}) from the known analytical solution to the Dirac Coulomb energy field of an infinitely-heavy point nucleus~\cite{1928-DC}.
This returns
\begin{equation}\label{eq:DiracPoint}
\Delta E_\text{DC-point}\equiv m_\mu(f-1)-E_0=-9.59~\text{eV},
\end{equation}
for $f\equiv \left[1+(Z\alpha)^2/\sqrt{1-(Z\alpha)^2})\right]^{-1/2}$, the Dirac factor for the $1S$ state.
%Eq.~(\ref{eq:DiracPoint} accounts for binding corrections  but not any finite size or vacuum polarization effects.

% FNS
In order to account for finite-size effects in terms of a perturbation series in $\rho$, we sum the one-, two-, and three-photon exchange contributions using parameters that are 
suitable for a Gaussian charge distribution, still within the 
approximation of an infinite nuclear mass.
The leading one-photon term is calculated to be
\begin{equation}\label{eq:FNS1}
\Delta E_\text{FNS1}\equiv -\frac{4}{3}E_0\,\rho^2=93.03~\text{eV}.
\end{equation}
It is model independent and proportional to $r_C^2$, which motivates focusing 
on this specific moment to begin with.

The next-to-leading order (NLO) term is due to two-photon exchange. It originates from Friar's work~\cite{1979-Friar}, and returns:
\begin{equation}\label{eq:FNS2}
\Delta E_\text{FNS2}\equiv\frac{2}{3}E_0\,(\rho\,r_Z/r_C)^3=-6.36~\text{eV},
\end{equation}
where $(r_Z/r_C)^3=32/(3 \sqrt{3\pi})$ for a Gaussian distribution~\cite{1979-Friar}.

The next-to-next-to-leading order (NNLO) term (due to three-photon exchange) 
is also model dependent and is the sum of two parts. 
The first part is a relativistic correction proportional to $\rho^2$ that returns (see Eq.~(68) of Ref.~\cite{2018-3P})
\begin{equation}\label{eq:FNS3-Rel.}
\Delta E_\text{FNS3}^\text{Rel.}
\equiv \Delta E_\text{FNS1} (Z\alpha)^2 \left[ 
\frac74-\gamma_e-\ln(2\,\rho\,r_{C2}/r_C)
\right]
=0.29~\text{eV}, % 0.293
\end{equation} 
where $\gamma_e\approx0.5772$ is the Euler-Mascheroni constant and $r_{C2}/r_C\approx1.014$ for a Gaussian distribution~\cite{2018-3P}.
%
%The order-of-magnitude of Eq.~(\ref{eq:FNS3-Rel.}) can be estimated as $E_0\,(Z\alpha)^2\,\rho^2\,\ln{\rho}\approx0.2\,$eV. 
The presence of the logarithmic term may be understood by expanding the 
relativistic correction to the finite size formula, which is proportional 
to $\rho^{\,2\,\gamma}-\rho^{\,2}\approx (Z\alpha)^2\rho^{\,2}\,\ln{\rho}$~\cite{1932-Racah}.
From the same logic, sub-leading relativistic corrections are of order $1\,$meV.
The second part of the NNLO finite-size correction is non-relativistic and proportional to $\rho^4$ (see Eq.~(68) of Ref.~\cite{2018-3P}). Here it returns 
\begin{align}\label{eq:FNS3-NR}
\Delta E_\text{FNS3}^\text{NR}
\equiv -\frac89 E_0\,\rho^4
 \left[
1+\gamma_e+\ln{(2\,\rho\,r_{C1}/r_C)}
+\frac{3}{20}(r_{CC}/r_C)^4\right]
=-0.12~\text{eV},
\end{align}
where $r_{C1}/r_C\approx0.559$ and $(r_{CC}/r_C)^4=5/3$ are for a Gaussian distribution.

The next terms in NRQED are unknown. Their magnitude can be roughly estimated 
from Fig.~(2) in Ref.~\cite{2026-Hybrid} to be $\approx20\,$meV. 
This demonstrates that a calculation perturbative in $\rho$ must be supplemented by an all-order (in $\rho$) calculation when going 
to slightly higher $Z$, depending on the desired accuracy goal in the radius extraction. 

%% eVP
Having accounted for the pure finite size effect to the desired level of accuracy, 
we turn to calculate the perturbation series related to eVP$_{11}$. Here,  
an infinitely heavy point nucleus is used. The leading term simply comes from a first-order 
perturbation theory (FOPT) calculation of the point electronic Uehling potential~\cite{1935-Uehling} 
with the infinitely-heavy point nucleus Schrödinger wavefunction.
The result is 
\begin{equation}\label{eq:eVP1-FOPT}
\Delta E_\text{eVP11}^\text{NR, FOPT}\equiv 2\,E_0 
\left(\frac{\alpha}{\pi}\right)\,F_{1s}(\kappa)=-138.95~\text{eV},
\end{equation}
where $F_{1S}(\kappa)$ can be found in Eq.~(1) of Ref.~\cite{2012-eVP11} for $\kappa\equiv 1/(a_0\,m_e)$.
%{\color{blue} [let us check if we define $a_0$ here, possibly with a numerical value]}
Because of the size of this correction, it is clear that second-order perturbation theory (SOPT) must be employed as well.
A calculation based on Eq.~(27) of Ref.~\cite{1996-Pachucki} returns 
\begin{equation}\label{eq:eVP1-SOPT}
\Delta E_\text{eVP11}^\text{NR, SOPT}=-0.32~\text{eV}.
\end{equation}
It has the expected order-of-magnitude: $E_0(\alpha/\pi)^2\approx-0.2\,$eV.
The third-order perturbation theory results in a negligible $1\,$meV~\cite{2021-1S-2S}.

Eq.~(\ref{eq:eVP1-FOPT}) and (\ref{eq:eVP1-SOPT}) are based on nonrelativistic wavefunctions. 
Due to the magnitude of the leading term, the leading relativistic correction to it 
must be accounted for. It is calculated (with an infinite nuclear mass) for the $1S$ state 
using Eq.~(25) of Ref.~\cite{2011-RelVP} and returns
\begin{equation}\label{eq:eVP1-Rel}
\Delta E_\text{eVP11}^\text{Rel.}\approx-0.12~\text{eV},
\end{equation}
whose magnitude is found to be consistent with the 
scaling relation $E_0\,(Z\alpha)^2\,\alpha/\pi\approx -0.1\,$eV.
Notice that in the infinite-nuclear mass limit, Eq.~(\ref{eq:eVP1-Rel}) has no nuclear spin dependence. 
The recoil correction to Eq.~(\ref{eq:eVP1-Rel}) is briefly discussed in subsection~\ref{sub:rad-rec}.
%Relativistic corrections to Eq.~(\ref{eq:eVP1-SOPT}) are smaller than $1\,$meV~\cite{2021-1S-2S}.

%% Mixed
In the fully perturbative approach (in $\rho$), the finite-size and eVP terms are calculated separately, so the mixed effect, which can be thought of as the finite-size correction to the eVP, must also be accounted for. 
The leading term does not depend on the nuclear charge distribution and may be calculated  from
\begin{equation}\label{eq:eVP1-FNS1}
\Delta E_\text{eVP11-FNS1}
\equiv \left( \frac{\alpha}{\pi}\right)\,\Delta E_\text{FNS1}\,C_1(1S, \kappa)=1.30~\text{eV}.
\end{equation}
$C_1(1S)=6.015$ is calculated by summing Eqs.~(14) and (15) of Ref.~\cite{2018-Saveli}. It slightly differs from the value given in Table~III of Ref.~\cite{2018-Saveli} because we use the full muon mass ({\em i.e.}, the non-recoil approximation).
The recoil correction to Eq.~(\ref{eq:eVP1-FNS1}) is considered separately in subsection~\ref{sub:rad-rec}.

The NLO term (in $\rho$) is calculated by summing Eqs.~(16) and (24) of Ref.~\cite{2018-Saveli} to be
\begin{equation}\label{eq:eVP1-FNS2}
\Delta E_\text{eVP11-FNS2}=-0.16\,\text{eV}.
\end{equation}
It is much larger than the naive estimation of 
$E_0\,\rho^3\,\alpha/\pi\approx-6\,$meV due to logarithmic enhancements 
(see Eqs.~(17) and (27) in~\cite{2018-Saveli}).
We note that Eq. (24) of Ref.~\cite{2018-Saveli} is suitable for a hard-sphere model (rather than the Gaussian model employed here). Nevertheless, the induced error is bounded by $11\,$eV, which is the difference between the calculation using an exponential model 
(see Eq.~(26) of Ref.~\cite{2018-Saveli}).

%% Result
Collecting all perturbative terms together, Eqs.~(\ref{eq:Bohr}) to (\ref{eq:eVP1-FNS2}), reproduces the all-order calculation result, Eq.~(\ref{eq:DCU}), to within $30\,$meV. A large part,
$21\,$meV, of this difference is due to missing elastic four-and-more-photon exchange corrections, 
and the remaining $9\,$eV is largely due to the implicit use of the hard-sphere model for the subleading mixed finite-size eVP contribution within the perturbative approach.

\subsection{Nonrelativistic Recoil}\label{sub:NRrec}
% Preamble

The calculations in subsection~\ref{sub:DCU} all assumed an infinitely-heavy nucleus and so employed the full (as opposed to reduced) muon mass.
We ascribe the phrase \textit{nonrelativistic recoil correction} to the difference between a calculated term when considering the nucleus to be of finite mass and the same term calculated with an infinite nuclear mass. 
%In non-relativistic mechanics, a recoil correction to any term may be calculated (to all orders in $\eta$) by taking the difference between the calculations with the reduced and full muon mass.
%This is not the case for a relativistic system, where the use of the reduced mass requires careful balancing by adding correction equations [Yennie].
Relativistic-recoil corrections are discussed in subsection~\ref{sec:rel-rec}.

% all order plus perturbative
To account for the nonrelativistic recoil to all orders in $\eta$ and $\rho$, we perform numerical calculations of the Dirac equation with a speed of light approaching infinity. We validated the numerical stability of this approach by reproducing the Bohr energy in the point nucleus limit.
Having taken this limit, we take the difference between the Schrödinger-Coulomb energy (that is, the Dirac-Coulomb energy in the limit of infinite light speed) calculated with and without a reduced mass. This results in the nonrelativistic, all order (in $\rho$ and $\eta$), recoil correction
\begin{equation}\label{eq:NR-Rec}
\Delta E_\text{Rec}^\text{NR}=556.50~\text{eV}.
\end{equation} 

We now reproduce the numerical value given in Eq.~(\ref{eq:NR-Rec}) with a series of calculations that are perturbative in $\rho$ (but all order in $\eta$). The point-nucleus term comes from scaling the Bohr energy with $m_r/m_\mu$ and subtracting Eq.~(\ref{eq:Bohr})
\begin{equation}\label{eq:NMS}
\Delta E_\text{Rec-point}^\text{NR}=\left[\frac{m_r}{m_\mu}-1\right]\,E_0=559.62~\text{eV}.
\end{equation}
The remaining parts are due to finite-size effects. 
As each of these is proportional to a higher power of $\rho$, they may be calculated simply by scaling with  $m_r/m_\mu$ to the appropriate power and taking the difference.
The leading-order nonrelativistic FNS recoil is thus
\begin{equation}\label{eq:FNS1-Rec}
\Delta E_\text{Rec-FNS1}^\text{NR}=\left[\left(\frac{m_r}{m_\mu}\right)^3-1\right]\,\Delta E_\text{FNS1}=-3.43~\text{eV}.
\end{equation}
The NLO FNS recoil is
\begin{equation}\label{eq:FNS2-Rec}
\Delta E_\text{Rec-FNS2}^\text{NR}=\left[\left(\frac{m_r}{m_\mu}\right)^4-1\right]\,\Delta E_\text{FNS2}=0.31~\text{eV}.
\end{equation} 
It is important to note that Eq.~(\ref{eq:FNS2-Rec}) is not the full recoil-correction at this order in $\rho$; it is only the nonrelativistic part~\cite{2025-RecFNS}. The residual relativistic-finite-size-recoil is discussed in subsection~\ref{sec:rel-rec}.
The same idea may be applied to the nonrelativistic part of the NNLO term [Eq.~(\ref{eq:FNS3-NR})], returning 
\begin{equation}\label{eq:FNS3-Rec}
\Delta E_\text{Rec-FNS3}^\text{NR}=
\left[\left(\frac{m_r}{m_\mu}\right)^5-1\right]\,\Delta E_\text{FNS3}^\text{NR}=0.01\,\text{eV}.
\end{equation}
The sum of Eqs.~(\ref{eq:NMS}) to (\ref{eq:FNS3-Rec}) reproduces Eq.~(\ref{eq:NR-Rec}) up to $\approx3\,$meV. 

%{\color{blue} Here is perhaps my most important concern, connected with Eqs.~(12.107) and (12.112) of Jentschura/Adkins. I will send an electronic version of the book with highlightings later.}

\subsection{Subleading Radiative Corrections}\label{sub:smaller}

In this subsection, we discuss radiative corrections other than eVP$_{11}$. These are small enough to be calculated in FOPT. To do so in the all-order (in $\rho$ and $Z\alpha$) approach, we use the Dirac-Coulomb wavefunctions and, in NRQED, the known 
equations that are based on FOPT with Schrödinger-Coulomb wavefunctions. 
Here again, the full muon mass (pertaining to an infinitely heavy nucleus) is employed. 
Nonrelativistic radiative-recoil corrections are dealt with separately in subsections~\ref{sub:NRRR} and \ref{sub:rad-rec}. 

%% KS
The first correction that we consider is the two-loop electronic vacuum polarization, often denoted as eVP$_{21}$ (standing for two loops, one vertex with the nucleus), which may be described by the Källén-Sabry potential~\cite{1952-KS}. Its magnitude, calculated numerically using the all-order approach 
(see, {\em e.g.} Ref.~\cite{1993-KS}), is 
\begin{equation}\label{eq:KS}
\Delta E_\text{eVP21}=-1.00~\text{eV}.
\end{equation}
The result
in Eq.~(\ref{eq:KS}) is all-order in both $Z\alpha$ and $\rho$. We could have included the potential directly in the self-consistent solution instead of solving perturbatively, or we could use the Dirac-Coulomb-Uehling wavefunctions in the perturbation calculation. 
However, the difference is found to be a few meV, which is commensurate with the various three-loop corrections that are neglected here.

Repeating the calculation in the nonrelativistic approximation, and with an infinitely-heavy point nucleus (we use Eq.~(8) from Ref.~\cite{2020-Repko}) returns
\begin{equation}\label{eq:KS-point}
\Delta E_\text{eVP21}^\text{NR, point}=-1.01~\text{eV}.
\end{equation}
Both results given in Eqs.~\eqref{eq:KS} and Eq.~\eqref{eq:KS-point} are a factor of 5 larger than the naive order-of-magnitude estimate $E_0\,(\alpha/\pi)^2\approx-0.2\,$eV.
The difference between Eqs.~(\ref{eq:KS}) and (\ref{eq:KS-point}) is $8\,$meV.
We trace this difference to the missing finite size correction to Eq.~(\ref{eq:KS-point}) in the perturbative calculation.

%% SE
We now consider the one-loop muon self-energy (SE1). In the all-order (in $\rho$ and $Z\alpha$) approach, we take the result from an upcoming publication~\cite{SE_FNS_Paul}
\begin{equation}\label{eq:SE1}
\Delta E_\text{SE1}=1.17~\text{eV}.
\end{equation}
Here, a hard-sphere distribution has been employed.
The leading-order perturbative calculation using an infinite nuclear mass returns
\begin{equation}\label{eq:SE1-point}
\Delta E_\text{SE1}^\text{NR, point}
\equiv-\frac{8}{3}\,E_0\,
\left(\frac{\alpha}{\pi}\right) \, (Z\alpha)^2\,
\left[\frac{5}{6}-\ln{k_0}(1S)-2\ln(Z\alpha)\right]
=1.17~\text{eV},
\end{equation}
where $\ln{k_0}(1S)\approx2.984$ is the relevant Bethe logarithm.
Although Eqs.~(\ref{eq:SE1}) and (\ref{eq:SE1-point}) are numerically identical, 
it does not mean that there are no missing terms. 
The leading correction to the one-loop self-energy due to two-Coulomb-photon 
exchange diagrams is well known, and in the infinite-nuclear-mass limit, it returns
\begin{equation}\label{eq:SE1-Rel}
\Delta E_\text{SE1}^\text{Rel., point}
\equiv -\,E_0\,\alpha\,(Z\alpha)^3\, 
\left[\frac{139}{16}-4\ln{2}\right] =0.05~\text{eV}.
\end{equation}
%
%It is an order-of-magnitude larger than the naive expectation $-E_0\,(Z\alpha)^3\,\alpha/\pi\approx3\,$meV due to an extra $\pi$ factor (discussed in subsection 3.3.1 of Ref.~\cite{2007-Eides}).
The leading finite-size correction to SE1 is calculated using  Appendix~E of Ref.~\cite{2023-Theory} to be 
\begin{equation}\label{eq:SE1-FNS}
\Delta E_\text{SE1-FNS}^\text{NR}=-0.04~\text{eV}.
\end{equation}
The result in
Eq.~(\ref{eq:SE1-FNS}) was calculated using an exponential charge distribution model. 
The difference from a Gaussian model is bounded by the differences in contributions between the exponential and hard-sphere models, which Fig.~(9) in Ref.~\cite{2018-Saveli} suggests is a few percent, 
{\em i.e.}, numerically negligible (below the meV range).
We see that Eqs.~(\ref{eq:SE1-Rel}) and (\ref{eq:SE1-FNS}) nearly cancel each other, so that summing them with Eq.~(\ref{eq:SE1-point}) reproduces Eq.~(\ref{eq:SE1}) to within $\approx11\,$meV. The difference can be ascribed to the different charge distribution models employed in the two calculations and the numerical convergence of the all-order method.

We also calculate the leading muonic vacuum-polarization ($\mu$VP$_{11}$) by considering the 
muonic Uehling potential perturbatively with the Dirac-Coulomb wavefunctions. The result of the all-order (in $\rho$ and $Z\alpha$) numerical calculation is
\begin{equation}\label{eq:uVP}
\Delta E_{\mu \text{VP11}}
=-0.05~\text{eV}.
\end{equation}
For the NRQED calculation, we can simply scale (again, in the limit of
an infinite nuclear mass) the known equation for the 
eVP$_{11}$ contribution in electronic atoms 
(see, {\em e.g.}, Eq.~(3.12) in Ref.~\cite{2007-Eides}) to obtain
\begin{equation}\label{eq:uVPpoint}
\Delta E_\text{$\mu$VP11}^\text{NR, point}
\equiv\frac{8}{15}\,E_0\,
\left(\frac{\alpha}{\pi}\right)
(Z\alpha)^2\,
=-0.05~\text{eV}.
\end{equation}
Adding the relativistic correction of $2\,$meV (see Eq.~(3.39) in Ref.~\cite{2007-Eides}) and 
the finite-size correction of $4\,$meV (see Appendix~E of Ref.~\cite{2023-Theory}) 
reproduces Eq.~(\ref{eq:uVP}) to better than $5\,$meV.

Finally, there is a small correction from the hadronic vacuum polarization (hVP$_{11}$). 
It is calculated in the all-order approach using the 
methods described in Ref.~\cite{2022-Hadronic} 
and returns
\begin{equation}\label{eq:hVP}
\Delta E_\text{$h$VP11}=-0.03~\text{eV},
\end{equation}
which agrees with the leading perturbative 
calculation (see, {\em e.g.}, Eq.~(71) in Ref.~\cite{2023-Theory})
\begin{equation}\label{eq:hVP-point}
\Delta E_\text{$h$VP11}^\text{NR, point}\approx0.675(16)\,\Delta E_{\mu \text{VP11}}\,=-0.03~\text{eV}.
\end{equation}
The $3\,$meV difference between Eq.~(\ref{eq:hVP-point}) and Eq.~(\ref{eq:hVP}) can be 
traced to our neglect of the finite-size correction to Eq.~(\ref{eq:hVP-point}).

\subsection{Nonrelativistic Recoil Corrections to Radiative Effects}\label{sub:NRRR}

The main radiative corrections eVP$_{11}$, eVP$_{21}$, and SE1 are large enough that calculating their reduced-mass dependence, which we here refer to as the recoil correction, may be necessary to meet the accuracy goal of this work.
The first two are dealt with nonrelativistically to FOPT in this subsection, where we compare the two approaches.
Subsection~\ref{sub:rad-rec} addresses the small recoil correction to eVP$_{11}$ in SOPT and the recoil correction to SE1 within NRQED only.
A discussion on radiative-recoil corrections beyond the use of the reduced mass is also given there.

In the all-order (in $\rho$) approach, we take the Schrödinger--Coulomb wavefunction 
from the numerical solution obtained for the potential of the extended nucleus and perturb it with the eVP$_{11}$ potential (also for an extended nucleus). 
This is done once with the reduced muon mass and once with the full mass.
The results are subtracted to give the all-order (in $\eta$ and $\rho$) recoil correction
\begin{equation}\label{eq:eVP1-Rec}
\Delta E_\text{eVP11-Rec.}^\text{NR, FOPT}=2.89~\text{eV}.
\end{equation}
In order to reproduce Eq.~(\ref{eq:eVP1-Rec}) with the perturbative (in $\rho$) calculation, 
we repeat the calculations leading to Eqs.~(\ref{eq:eVP1-FOPT}) and (\ref{eq:eVP1-FNS1}) 
but now using the reduced mass and taking the difference, thus obtaining
\begin{equation}\label{eq:eVP1-Rec-NR}
\Delta E_\text{eVP11-Rec.}^\text{NR, point}
\equiv2\,E_0 \left(\frac{\alpha}{\pi}\right)\,
\left[\frac{m_r}{m_\mu}\,F_{1s}\left(\frac{m_r}{m_\mu}\kappa\right)-F_{1s}(\kappa)\right]
=2.94~\text{eV}
\end{equation}
and
\begin{equation}\label{eq:eVP1-Rec-FNS}
\Delta E_\text{eVP11-FNS1-Rec.}^\text{NR}
\equiv \left(\frac{\alpha}{\pi}\right)\,\Delta E_\text{FNS1}
\left[\left(\frac{m_r}{m_\mu}\right)^3\,C_1\left(1S,\frac{m_r}{m_\mu}\kappa\right)-C_1(1S, \kappa)\right]
=-0.05~\text{eV}.
\end{equation}
The sum of Eqs.~(\ref{eq:eVP1-Rec-NR}) and (\ref{eq:eVP1-Rec-FNS}) 
reproduces Eq.~(\ref{eq:eVP1-Rec}) to $10\,$meV. 

For eVP$_{21}$, an analogous all-order (in $\rho$) calculation with the reduced mass and the following subtraction of Eq.~(\ref{eq:KS}) returns
\begin{equation}\label{eq:eVP2-Rec}
\Delta E_\text{eVP21-Rec.}^\text{NR}=0.02~\text{eV} \,.
\end{equation}
In order to reproduce Eq.~(\ref{eq:eVP2-Rec}) with NRQED, we repeat the calculations leading to Eq.~(\ref{eq:KS-point}) using the reduced mass and subtract it, thereby obtaining
\begin{equation}\label{eq:eVP21-Rec-point}
\Delta E_\text{eVP21-Rec.}^\text{NR, point}=0.02~\text{eV}.
\end{equation}
The two agree to better than $1\,$meV.

\subsection{Theoretical Predictions and Uncertainty}
\label{sub:rec}

The sum of all all-order (in $\rho$) calculations; given in Eqs.~(\ref{eq:DCU}), (\ref{eq:NR-Rec}), (\ref{eq:KS}), (\ref{eq:SE1}), 
(\ref{eq:uVP}), (\ref{eq:hVP}), (\ref{eq:eVP1-Rec}), and (\ref{eq:eVP2-Rec}); is
\begin{equation}\label{eq:sumAO}
E_{1S}^\text{AO}=-44\,513.08(2)~\text{eV}.
\end{equation}
Our $20\,$meV uncertainty corresponds to the sum of differences between all-order (in $\rho$) and the NRQED calculations that cannot be simply traced to a missing elastic 4-photon-exchange correction. It is attributed to numerical convergence and residual model dependence. 
Eq.~(\ref{eq:sumAO}) is not yet our final recommended value for the ground-state 
binding energy, as it is missing small corrections that we discuss next.

\section{Additions from NRQED}\label{sec:add}
Here we collect small corrections that have not been addressed in section~\ref{sec:main}. 
Given the uncertainty of $20\,$meV in Eq.~(\ref{eq:sumAO}), we choose to consider contributions that are larger than $10\,$meV. 
To validate the calculations in this section, we reproduce the values in Table I of Ref.~\cite{2023-Theory} for $^1$H, $^2$H, $^3$He, and $^4$He to all of the digits given there.
\subsection{Negligible Higher--Order Radiative Corrections}
Corrections that are even smaller than our threshold for inclusion are given in Ref.~\cite{2021-1S-2S}. 
These are the higher-order (in Coulomb vertices) Wichmann-Kroll 
correction ($\approx8\,$meV), 
the higher-order light-by-light scattering correction
($\approx-2\,$meV), the irreducible three-loop eVP correction in FOPT 
($\approx-4\,$meV), and further 
reducible three-loop eVP diagrams ($\approx-5\,$meV).
These add up to a sum of about $-3\,$meV, which is completely negligible.

\subsection{Relativistic Recoil}\label{sec:rel-rec}

In subsection~\ref{sub:NRrec}, we have calculated non-radiative recoil corrections in the nonrelativistic approximation. 
One could imagine that relativistic recoil corrections would contribute at the 
level of $-\eta\,E_0(\alpha Z)^2\approx0.5\,$eV, which is definitely 
non-negligible. However, in the point-nucleus approximation, the leading  
term (in $\eta$) cancels for $1S$ states (see, {\em e.g.}, Eq.~(16) of Ref.~\cite{2022-two-body}), 
so the nonrelativistic recoil approximation is much better than expected, and the subleading order starts at $(Z\alpha)^3$ instead of $(Z\alpha)^2$.

For a point nucleus, the all-order (in $Z\alpha$), leading order (in $\eta$), relativistic correction to the recoil is given by
\begin{equation}\label{eq:sal}
\Delta E_\text{Rec.}^\text{Rel., point}
\equiv -\frac2\pi\eta\,E_0\, (Z\alpha)^3\,P(Z)
=0.04~\text{eV},
\end{equation}
with $P(Z=4)\approx4.45\,$ interpolated from table I of Ref.~\cite{1995-Shabaev}.
This result is well approximated (to within $1\,$meV) by the leading order (in $Z\alpha$) term first calculated by Salpeter~\cite{1952-Salpeter}.

%Focusing first on relativistic recoil corrections that are leading-order 
%in $\eta$, the largest missing term is that of the relativistic recoil correction of order $\eta (Z\alpha)^3 \, E_0$,
%which was calculated by Salpeter~\cite{1952-Salpeter} for the ground state, 
%and which returns
%
%\begin{equation}\label{eq:sal}
%\Delta E_\text{Rec.}^\text{Rel., point}=0.04~\text{eV}.
%\end{equation}
%
%Its size comes from a logarithmic enhancement that makes its natural order-of-magnitude 
%$-\eta\,E_0(\alpha Z)^3\ln(Z\alpha)\approx0.04\,$eV.
%The next order (in $Z\alpha$) point recoil term may be calculated using, 
%{\em e.g.}, Eq.~(5) from Ref.~\cite{2023-Adkins} to be a negligible $-0.5\,$meV. A similar all order (in $Z\alpha$) calculation based on table I of Ref.~\cite{1995-Shabaev} gives a residual of $-0.6\,$meV in agreement with the perturbative calculation.

In light of the size of Eq.~(\ref{eq:sal}), it may be necessary to account for the 
leading relativistic correction to the recoil correction for the finite size effect. We estimate this effect
based on the difference in the nonrelativistic reduced-mass shift of the
two-photon exchange term
given in Eq.~(\ref{eq:FNS2-Rec}) 
and the full calculation provided by Eq.~(49) of the recent work~\cite{2025-RecFNS}.
The result is a negligible $4\,$meV.
Similarly, Eq.~(\ref{eq:FNS3-Rec}) differs from a calculation based on 
Eq.~(71) of Ref.~\cite{2025-RecFNS} by less than $1\,$meV.
Thus, for this system, relativistic corrections to the finite-size recoil corrections may safely be neglected.

In subsection~\ref{sub:NRrec}, the nonrelativistic recoil was calculated to 
all orders in $\eta$ through the use of the reduced mass. 
We have just seen that the linear (in $\eta$) leading relativistic recoil vanishes for a point nucleus. This is true irrespective of the nuclear spin.
However, the quadratic (in $\eta$) relativistic recoil correction does not 
necessarily disappear. It depends on the relativistic equation of motion 
that the nucleus obeys, which depends on its spin, as well as the fine details of the definition of the 
charge radius~\cite{jentschura2011proton,  adkins2026relativistic}.
If the heavy partner was a spin $1/2$ point particle, such as a positive muon, the correction is simply $-0.5\,\eta^2\,E_0\,(\alpha Z)^2\approx4\,$meV~\cite{barker1955reduction}. 
For $^9$Be, whose nuclear spin is $3/2$, there is no established definition of the radius. Nevertheless, as the ambiguity induced in the calculated ground-state energy is only of the order of $4\,$meV, so that we can choose to defer this discussion to future work.

\subsection{Added Radiative--Recoil Corrections}\label{sub:rad-rec}
In subsection~\ref{sub:NRRR}, we compared the nonrelativistic 
recoil corrections to eVP$_{11}$ in FOPT. 
Here, we supplement this with the nonrelativistic recoil corrections to 
eVP$_{11}$ in SOPT by subtracting the result from Eq.~(\ref{eq:eVP1-SOPT}), which was originally obtained by employing wave functions
for an infinitely heavy nucleus, from an analogous 
calculation that employs the reduced mass. The difference is
\begin{equation}\label{Eq:eVP-SOPT-Rec}
\Delta E_\text{Rec.-eVP11-SOPT}^\text{NR., point}=0.01~\text{eV}.
\end{equation}
Similarly, subtracting the result given in Eq.~(\ref{eq:eVP1-Rel}) for the relativistic corrections to the one-loop eVP, as obtained using an infinite nuclear mass, from an analogous calculation 
that employs the reduced mass returns the leading order (in $\eta$ and $Z\alpha$) 
relativistic recoil to eVP$_{11}$. 
It amounts to a numerical shift of $-2\,$meV and can thus be neglected.

The all-order (in $\eta$) recoil correction to the leading self energy correction is obtained by 
reintroducing the reduced mass and subtracting the non-recoil result [Eq.~(\ref{eq:SE1-point})],
giving 
\begin{align}\label{eq:SE-Rec}
\Delta E_\text{SE1-Rec.}^\text{NR., point}
\equiv \left[\left(\frac{m_r}{m_\mu}\right)^3-1\right]
\Delta E_\text{SE1}^\text{NR, point}+
\frac{8}{3}\,E_0\,\left( \frac{\alpha}{\pi}\right)\,(Z\alpha)^2\,
\left(\frac{m_r}{m_\mu}\right)^3\ln{\left(\frac{m_r}{m_\mu}\right)}
\nonumber\\
=-0.05~\text{eV}.
\end{align}
The subleading (in $Z\alpha$) recoil correction to the point-nucleus self energy can be estimated from Eq.~(5.2) of Ref.~\cite{2007-Eides} to be under $1\,$meV.

The nuclear self-energy contribution vanishes for an infinitely heavy nucleus, 
so it can be thought of as a quadratic (in $\eta$) radiative-recoil correction.
Although it can be estimated using Eq.~(41) of Ref.~\cite{Codata2018} to be $5\,$meV, 
its definition is not free from ambiguity for a nucleus of finite size and 
spin-$3/2$ (see, {\em e.g.}, the discussion in subsection 5.1.3 
of Ref.~\cite{2007-Eides}). Nevertheless, the smallness of this correction 
allows us to defer the discussion to future work, as we did for the relativistic quadratic recoil.

\subsection{Self--Energy and Vacuum Polarization (SE-eVP)}

In subsection~\ref{sub:smaller}, we have calculated the largest two-loop QED contribution, namely eVP$_{21}$.
Consulting Fig.~2 of Ref.~\cite{2024-SEVP}, the largest two-loop calculation beyond eVP$_{21}$ is the eVP correction to the SE. It can be calculated directly from eq.~(31) in Ref.~\cite{2024-SEVP} to be 
\begin{equation}\label{eq:SE-VP}
\Delta E_\text{eVP11-SE1}^\text{NR, point}=0.01~\text{eV}.
\end{equation}

\subsection{Recommended Value with Uncertainty}\label{sub:sumadd}

The sum of the added radiative effects, from Eqs.~(\ref{eq:sal}) to (\ref{eq:SE-VP}), 
amounts to $14(6)$\,meV, where the uncertainty stems from ambiguities related 
to the interplay between relativistic quadratic 
recoil, nuclear self energy, and the definition of the charge radius. 
Adding this to Eq.~(\ref{eq:sumAO}) gives our recommended value for the binding energy
\begin{equation}\label{eq:rec}
E_{1S}=-44\,513.07(2)\,\text{eV}.
\end{equation}

\subsection{Binding Energies of $P$ States}\label{sub:p}

In gross structure measurements of light muonic atoms, the main transition used for radius 
determinations is $2P$--$1S$ (see Ref.~\cite{2004-Fricke}). Because the fine (and hyperfine) structure is rarely distinguished by the experimental resolution, it is often useful to treat the $P$-level multiplet as one level whose energy lies in the center of gravity.
In light systems, the cascade process is expected to populate fine-structure components 
statistically; {\em i.e.}, according to $2J+1$, so that the center of gravity takes 
the form $\overline E_{2P}\equiv(1/3)E_{2P_{1/2}}+(2/3)E_{2P_{3/2}}$.
The hyperfine structure may be dealt with in a similar manner.

We repeat the NRQED calculations for the $P$-state energies. This is much simpler than for the 
$1S$ level because there are only a few terms that are larger than $2\,$meV. 
Specifically, these are the ones that lead to Eqs.~(\ref{eq:Bohr}), (\ref{eq:DiracPoint}), 
(\ref{eq:eVP1-FOPT}), (\ref{eq:eVP1-Rel}), (\ref{eq:NMS}), and (\ref{eq:KS}). Additionally, the 
leading-order relativistic recoil [see, {\em e.g.}, Eq.~(16) in~\cite{2022-two-body}), which is zero for the $1S$ state, increases the center-of-gravity energy for the $2P$ states by $8\,$eV and must be included.
The resulting center-of-gravity energy is 
\begin{equation}\label{eq:pCOG}
\overline E_{2P}=-11\,120.52\,\text{eV}.
\end{equation}
It is independent of finite-size and other nuclear-structure effects at the targeted level of accuracy.

\section{Parametrization}\label{sec:par}

In this section, we derive a simple equation that enables the community to 
update $r_C$ based on the results of ongoing and future muonic-$^9$Be spectroscopy experiments. 
In order to simplify the equation as much as possible, we limit its applicability range to $|r_C-2.519\,\text{fm}|<5\%$.
This range is an order-of-magnitude larger than the statistical accuracy in the reference radius~\cite{1972-Be}, so it is expected to cover any reasonable deviations from it.

First, we note that the elastic three-photon-exchange contribution [the sum of Eqs.~(\ref{eq:FNS3-Rel.}) and (\ref{eq:FNS3-NR})] changes only by a negligible 6\,meV for the $5\%$ variation in the radius. This is due to a partial cancellation between the relativistic 
[Eq.~(\ref{eq:FNS3-Rel.})] and nonrelativistic [Eq.~(\ref{eq:FNS3-NR})] parts.
This fact, which may not be true for other systems, enables a rather simple parametrization around the reference radius
\begin{equation}\label{eq:par}
E_{1S}(r_C)-E_{1S}(2.519\,{\rm fm}) = F_1\left[r_C^2-(2.519\,{\rm fm})^2\right]+F_2\left[r_C^3-(2.519\,{\rm fm})^3\right],
\end{equation}
that is valid to within $10\,$meV in the range $r_C=2.39-2.65\,$fm, with the constant coefficients $F_1$ and $F_2$ being
determined next.

The leading $F_1$ coefficient is calculated by 
dividing the sum of the largest terms that have a quadratic dependence on 
$r_C$ [Eqs.~(\ref{eq:FNS1}), (\ref{eq:eVP1-FNS1}), and (\ref{eq:FNS1-Rec})] by $r_C^2$. The returned value is
\begin{equation}\label{eq:R1}
F_1
%=(93.03+1.29-3.43)/2.519^2
=14.32\,\text{eV/fm}^2.
\end{equation}
Similarly, the subleading $F_2$ coefficient is calculated by summing Eqs.~(\ref{eq:FNS2}), (\ref{eq:FNS2-Rec}), and (\ref{eq:eVP1-FNS2}), 
and dividing by $r_C^3$, which returns
\begin{equation}\label{eq:R2}
F_2
%=(-6.36+0.31-0.16)/2.519^3
=-0.389\,\text{eV/fm}^3.
\end{equation}

The final parameterization of the energy, valid in the range $r_C=2.39-2.65\,$fm, is obtained by subtracting Eqs.~(\ref{eq:rec}) and~(\ref{eq:par}) from Eq.~(\ref{eq:pCOG}), resulting in
\begin{equation}\label{eq:final}
E_{2P-1S}(r_C)=33\,392.55(2)-14.32\,(r_C^2-2.519^2)\,+0.389\,(r_C^3-2.519^3)\,+\Delta E_\text{NS}\,,
\end{equation}
where, in order to simplify the notation, we indicate energies measured in eV and radii in fm. 
Notice that we introduced $\Delta E_\text{NS}$, 
which denotes the sum of corrections to the $2P$--$1S$ transition energy that depends on the nuclear model and is not covered by this work. 
It is approximately $1\,$eV~\cite{1985-Drake, 2026-Guide}. 
In light nuclei ($Z\leq8$), the uncertainty of $\Delta E_\text{NS}$ is dominated by that of the nuclear polarization (see, {\em e.g.}, Fig.~4 of Ref.~\cite{2025-Mirror}). Considering that recent calculations of nuclear polarization contributions to muonic atoms quote uncertainties of $10$-$20\%$~\cite{2022-NP, beyer2025modern,2025-Pb,2026-Guide, beyer2025relativistic, 2026-Hybrid}, 
{\em i.e.}, $0.1$-$0.2\,$eV for muonic $^9$Be,
the pure QED uncertainty of $20\,$meV that we quote in Eq.~(\ref{eq:final}) is not expected to limit radius extractions in this system for years to come.

\section{Conclusions}

Over the last few decades, two different approaches have been used for the evaluation of theoretical 
predictions for bound muonic systems over the ranges of low and intermediate nuclear charge numbers $Z$.
For high $Z$, it was recognized that the nuclear-size effect (even details thereof, connected with the 
shape of the nuclear charge distribution) should be included to 
all orders right from the start. For low nuclear charge numbers,
it was recognized that the larger values of the recoil corrections encountered in the 
calculations demanded an approach that carefully considers the dependence of the 
bound-state energies beyond the non-recoil limit of an infinite nuclear mass.

Here, we demonstrate that, with careful bookkeeping, both the fully perturbative approach (in the nuclear-size effect) commonly employed for the lightest system  and the alternative approach that is nonperturbative in the finite size and, for the most part, even nonperturbative in $Z\alpha$,
but applied typically in the non-recoil limit, agree to better than one part per million. 
The agreement found here gives us confidence in the numerical stability of the all-order calculations. Furthermore, we hope that our work will serve as a guide for the bookkeeping of the various recoil corrections that must supplement relativistic calculations in medium-Z muonic atoms.

Another result of this work is a well-tested parametrization of the muonic $^9$Be $2P$--$1S$ transition energy that was recently measured at PSI in terms of the charge radius and nuclear-structure corrections. 
%{\color{blue} [maybe, to add a citation, even if the work has not yet been published. maybe some ``private communication'' or so]}

%Our aim is to validate both approaches and build a connection between the respective communities while placing particular emphasis on the transparency and reproducibility of the results.

\section*{Acknowledgmemts}

We thank O. Eizenberg and K. Pachucki for giving valuable corrections.
B.O is thankful for the support of the Council for Higher Education Program for Hiring Outstanding Faculty Members in Quantum Science and Technology. S. R. gratefully acknowledges the Technion and the Davis Postdoctoral Fellowship.
Big thanks goes out to Randolf Pohl for inspiring this work.
The work of U.D.J.~was supported by the National Science Foundation 
(grant PHY--2513220).

%%%%%%%%%%%%%%%%%%%%%%%%%%%%%%%%%%%%%%%%%%

\end{document}